\documentclass[conference]{IEEEtran}

\usepackage{url}

\usepackage{breakurl}
\usepackage[hidelinks]{hyperref}

\usepackage{graphicx}
\usepackage[dvipsnames]{xcolor}
\usepackage{amsmath}
\usepackage{tikz}
\usepackage{booktabs}

\usepackage{listings}
\definecolor{codegreen}{rgb}{0,0.6,0}
\definecolor{codegray}{rgb}{0.5,0.5,0.5}
\definecolor{codepurple}{rgb}{0.58,0,0.82}
\definecolor{backcolour}{rgb}{0.95,0.95,0.92}
\lstdefinestyle{mystyle}{
    backgroundcolor=\color{backcolour},   
    commentstyle=\color{codegreen},
    keywordstyle=\color{magenta},
    numberstyle=\tiny\color{codegray},
    stringstyle=\color{codepurple},
    basicstyle=\ttfamily\footnotesize,
    breakatwhitespace=false,         
    breaklines=true,                 
    captionpos=b,                    
    keepspaces=true,                 
    numbers=left,                    
    numbersep=5pt,                  
    showspaces=false,                
    showstringspaces=false,
    showtabs=false,                  
    tabsize=2
}
\lstdefinelanguage{yaml}{
    basicstyle=\color{BlueViolet}\fontfamily{lmss}\selectfont\small,
    rulecolor=\color{black},
    sensitive=false, 
    string=[s]{'}{'},
    stringstyle=\color{BlueViolet},
    morestring=[b]{"}, 
    columns=fullflexible,
    keywords={true,false,null,y,n},
    keywordstyle=\color{PineGreen},
    keywordstyle=[2]\color{BlueViolet}, 
    numbers=none,
    showstringspaces=false,
    breaklines=true,
    frame=top bottom,
    comment=[l]{:},
    morecomment=[s]{/*}{*/},
    commentstyle=\color{RubineRed}
}
\usepackage{subcaption}
\usepackage{caption}
\DeclareCaptionFont{black}{ \color{black} }
\DeclareCaptionFormat{listing}{
  \parbox{\textwidth}{\hspace{15pt}#1:#3}
}
\usepackage[section]{placeins}

\usepackage{cite}
\usepackage{amsmath,amssymb,amsfonts}
\usepackage{textcomp}
\def\BibTeX{{\rm B\kern-.05em{\sc i\kern-.025em b}\kern-.08em
    T\kern-.1667em\lower.7ex\hbox{E}\kern-.125emX}}

\newcommand\copyrighttext{
  \footnotesize \textcopyright 2025 IEEE. Personal use of this material is permitted. Permission from IEEE must be obtained for all other uses, in any current or future media, including reprinting/republishing this material for advertising or promotional purposes, creating new collective works, for resale or redistribution to servers or lists, or reuse of any copyrighted component of this work in other works. DOI: \href{https://doi.org/10.1109/NetSoft64993.2025.11080536}{10.1109/NetSoft64993.2025.11080536}}
\newcommand\copyrightnotice{
\begin{tikzpicture}[remember picture,overlay]
\node[anchor=south,yshift=10pt] at (current page.south) {\fbox{\parbox{\dimexpr\textwidth-\fboxsep-\fboxrule\relax}{\copyrighttext}}};
\end{tikzpicture}%
}

\begin{document}

\title{A VPN-as-a-Service Tailored Enabler for Computing-constrained Environments}

\author{
\IEEEauthorblockN{
Carolina Fern\'{a}ndez-Mart\'{i}nez\IEEEauthorrefmark{1}\IEEEauthorrefmark{2},
C\'{e}sar Cajas Parra\IEEEauthorrefmark{1},
Shuaib Siddiqui\IEEEauthorrefmark{1}
}
\IEEEauthorblockA{
    \IEEEauthorrefmark{1}i2CAT Foundation, \mbox{08034 Barcelona, Spain}\\
    \{carolina.fernandez, cesar.cajas, shuaib.siddiqui\}@i2cat.net\\
}
}

\maketitle
\copyrightnotice

\begin{abstract}
Industry has embraced Zero Trust (ZT) architectural tenets and implementations for cloud-native environments, following stricter security requirements to both internal and external tenants.
Among others, these approaches combine fine-grained identity management and monitoring for both inventorying and better analysing the devices' security posture for overall protection, along with strict separation of concerns and isolation to enforce minimal privilege.
Networking-wise, ZT approaches rely as well on isolation and least privilege; enacted by separate, secure tunnels per tenant connecting to a given infrastructure. Such implementations can also be applied to the connectivity within and towards experimental infrastructures.
In this sense, this work contributes the design and evaluation of a cloud-native VPN-as-a-Service (VPNaaS) that can be (i) easily orchestrated to deploy on-the-fly, separate tunnels per each tenant remotely connecting to the infrastructure; (ii) integrated with common Identity and Access Management (IAM) tools, key to ZT deployments; and (iii) adapt to computing- or entropy- constrained environments. This solution is customisable and allows, among others, to select from RSA or Elliptic Curves (EC) as key generation algorithm and their parameters to achieve more secure keys and adapt to resource-constrained environments.

\end{abstract}


\IEEEpeerreviewmaketitle

\section{Introduction}
\label{sec:introduction}

In today's IT-centric organisations, reliance on virtualised services and remote access is now a key component and reaches very different sectors, from those of high criticality, including the public administration, to the myriad of Small and Medium-sized Enterprises (SMEs) organisations account for most of the market.
Subsequently, several supranational\footnote{https://www.insideprivacy.com/covid-19/guidance-released-by-eu-authorities-on-how-to-ensure-it-security-when-working-remotely/} and and national institutions, like the European Commission and Data Protection Agencies (DPAs) in Europe or the National Institute of Standards and Technology (NIST) in the; have US published both regulations \cite{eu_directive_2022_2025}, standards and recommendations \cite{Souppaya_Scarfone_2016} for almost a decade now to guide on how to increase security protection in networks and devices.
On the one hand, EU regulatory documents are transcribed by each Member State's Computer Security Incident Response Team (CSIRT) so to define specific security measures and controls to be enforced by critical (e.g. public) sectors \cite{ie_ncsc_nis2_quick_guide}.
On the other hand, US guidance documents help setting up secure remote connectivity towards organisations' infrastructure and services for different remote working scenarios \cite{Souppaya_Scarfone_2016}: (i) IPSec and Secure Socket Layer (SSL)-based tunnels; (ii) access portals; (iii) remote desktops and (iv) direct access to applications.

Whilst earlier reports already introduced the lack of trust in the networks connecting towards some infrastructure, the area of a trusted infrastructure was usually delimited by the perimeter of its organisation.
With the publication of BeyondCorp \cite{Ward_Beyer_2014} and the latter standardisation of the Zero Trust (ZT) architecture by NIST \cite{Rose_Borchert_Mitchell_Connelly_2020} and its wider adoption by organisations, the internal networks were also considered subject to cyber threats, and every segment was deemed untrusted by default.
In a nutshell, ZT combines fine-grained identity management and network segmentation along with monitoring and observability to enforce security more stringently.
Thus, when implementing ZT in an infrastructure that allows remote access and offers multiple services, it is crucial to adopt (i) identity management and access control mechanisms with adequate interoperability across multiple offered services and implement well-known authentication and authorisation protocols in order to ensure data confidentiality; and (ii) network segmentation across multiple tenants or users, ensuring isolated flow transmission per connection that is integrated with the identity management.

This work proposes a cloud-native L3 Virtual Private Network-as-a-Service (VPNaaS) solution for (i) zero-touch deployments in common data centres via its Helm chart; extending OpenVPN (ii) to support EC curves and easily tuning the key generation algorithm and parameters (e.g. key size, curve) at boot time to adapt to security requirements and computational constraints.
It also delivers (iii) custom authentication logic in two ways (tokens and \textit{basicauth}) and integrates with the well-known Keycloak Identity and Access Management (IAM) to (iv) support Single Sign-On (SSO) across the 6G-BRICKS\footnote{https://6g-bricks.eu} project's experimental infrastructures, ensuring that end-users (i.e. experimenters) connect in a secure and isolated fashion.
Altogether, this aims at reducing the complexity of integrating with existing stacks, preserving reduced deployment times and suitable security.

The rest of the paper continues with a brief literature review (Section \ref{sec:related-work}), the design and implementation details on the proposed enabler for a ZT framework and its application in the project (Section \ref{sec:design-implemensation}, followed by the evaluation of its deployment (Section \ref{sec:validation}) and concluded by final remarks and future directions to evolve (Section \ref{sec:conclusions}).

\section{Related work}
\label{sec:related-work}

Previous literature analysed ZT requirements, adequacy and security stance of the organisations, as well as proposing ZT frameworks and implementations for specific use cases.

As part of the ZT analysis, \cite{Yeoh_Liu_Shore_Jiang_2023} defines a maturity assessment framework consisting of eight dimensions, based on the ZT principles, so that organisations could evaluate the maturity of their implementation.
From these dimensions, this work focuses on identity management (ID3-ID4), network (NE1-NE3), infrastructure (IN3) and orchestration (AO1).
The adequacy of the ZT Architecture (ZTA) is qualitatively evaluated in \cite{Fernandez_Brazhuk_2024} regarding performance and feasibility, along as suitability to protect against threats and to implement. Among others, it points at the potential check and policy explosion and large overhead due to possibly too detailed access controls.
In \cite{Itodo_Ozer_2024}, academic and grey literature is grouped into four big blocks, encompassing aspects of the ZT core tenets and concluding that works tend to focus on very specific categories rather than providing holistic views. This is consistent with ZT acting as an umbrella and integrator of multiple techniques.

Some of these dimensions and aspects are now highlighted.
On identity management, \cite{ida_simpson_foltz_2021} applies ZT to federated enterprises, using strong Public Key Infrastructure (PKI) and traditional X.509 certificates for authentication, binding these to the authorisation claims that use the Security Assertion Markup Language (SAML); or alternatively resorting to weaker, non-PKI identities.
Inspired by the PKI root of trust, \cite{Poirrier_Cailleux_Clausen_2023} introduces Remote ATtestation procedureS (RATS) to prove trust across domains during the process to access request to resources and to exchange the requester's identity attributes by remotely verifying their integrity.
For network segmentation, 
\cite{Zohaib_Sajjad_Iqbal_Yousaf_Haseeb_Muhammad_2024} proposes a general, theoretical ZT-VPN framework with OpenVPN tunnels for the ZT Policy Enforcement Point (PEP), which performs identity management before granting access using both a certificate with basic authentication on the Identity Enforcement Point.

Moving to specific ZT implementations based on the Software-Defined Perimeter (SDP) \cite{csa_sdp_v20} framework, \cite{Abdelhay_Bello_Refaey_2024} connect mobile core components securely and reliably using dynamic firewalls and SDP-based authentication and authorisation. It also compares initialisation times, CPU and memory consumption and vulnerability surface between VPN and SDP.
Finally, ad-hoc L3 VPN deployments are more common and span from earlier Software-Defined Networking (SDN) dynamically configuring Multiprotocol Label Switching (MPLS) in core network routers \cite{Van_Der_Pol_Gijsen_Zuraniewski_Romao_Kaat_2016} to zero-touch Network Functions Virtualisation (NFV) based orchestration of meshed WireGuard instances \cite{Direito_Gomes_Gomes_Aguiar_2023}.

The proposed work delivers a ZT-compliant VPN implementation, designed to be integrated with SDP to provide a more balanced network segmentation that limits the number of checks and policies required in micro-segmentation setups \cite{Fernandez_Brazhuk_2024} by reducing the policy scope to reachable clients. This results in an on-demand and cloud-native tunnel setup, compared to some prior works like \cite{Van_Der_Pol_Gijsen_Zuraniewski_Romao_Kaat_2016} and \cite{Direito_Gomes_Gomes_Aguiar_2023}, respectively. It relies on open-source solutions, running common authentication enforced by major open-source IAM and seamlessly supporting SSO; as well as allowing custom tunnel configuration to adjust to a desired security level and computation constraints for generated keys, paving the way to address future steps from other works, e.g.\cite{Zohaib_Sajjad_Iqbal_Yousaf_Haseeb_Muhammad_2024}.
\section{Design \& implementation}
\label{sec:design-implemensation}

This VPNaaS solution is one of the ZT Security Mechanisms (ZTSM), along with a SDP Gateway; both operating in the Security Data Plane.
These are part of the 6G-BRICKS ZT Security Framework (ZTSF), where connections must (i) support seamless authentication process across all platform services by using SSO; and (ii) provide network segmentation by enabling secure, isolated access per experimenter to the experimental testbeds, in particular to perform access control towards the protected back-end services and/or endpoints.

\subsection{Design}

\begin{figure}[tb]
    \centering
    \includegraphics[width=0.5\textwidth]{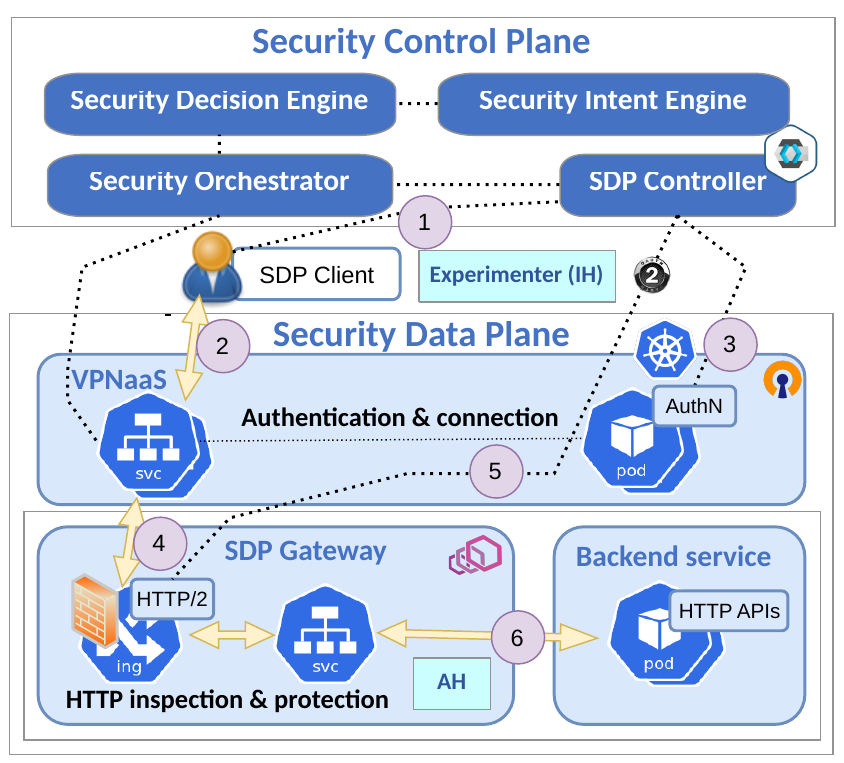}
    \caption{VPNaaS interacting with SDP within the ZTSF}
    \label{fig:use-case:integrated-view}
\end{figure}

The following requirements were first elicited for the VPNaaS ZTSM: (i) enact authenticated access for experimenters before establishing a VPN tunnel, adopting JSON Web Tokens (JWT) that are managed by the central IAM to offer SSO authentication, as well as basic authentication; (ii) offer a tunnel per user, where each experimenter is assigned a dedicated VPN server with different configuration parameters and keys, to ensure isolated access to the protected back-end services running in the experimental testbed; (iii) be compliant with Kubernetes (K8s) cloud-native orchestration, using Helm charts; and (iv) offer automated tunnel setup, server configuration and client profile generation in a zero-touch manner that allows later configuration retrieval by the orchestrator.
Besides, as tests graduated from pre-production to production environments with different types K8s clusters and hardware, computational constraints were found that impacted the 60 seconds threshold expected for the end-to-end security establishment in the project \cite{6gbricks_d23}.
Given that, a new requirement was introduced to (v) allow configuration before deployment of the key generation algorithm that takes place before setting up the tunnel, to either use RSA or Elliptic Curve Digital Signature Algorithm (ECDSA) keys.

Prior to design, OpenVPN and WireGuard were evaluated based on the criteria of performance, configurability, security, and integration with modern authentication mechanisms.
It is worth noting that, while WireGuard is known for its higher throughput, simpler configuration and larger community nowadays, with external works on e.g. authentication\footnote{https://github.com/Place1/wg-access-server}; OpenVPN seemed to offer better versatility, fine-grained control over encryption and routing, and built-in, robust support for authentication mechanisms (e.g. LDAP, RADIUS, SAML).
Specifically, its plug-in system\footnote{https://community.openvpn.net/openvpn/wiki/PluginOverview} supports scripts as hooks, allowing straightforward integration with external authentication (here, the project's IAM tool - Keycloak) and enabling SSO authentication.
OpenVPN was consequently selected over WireGuard; and in particular, Community Edition (CE) was chosen instead of OpenVPN Access Server (AS) and CloudConnexa\footnote{https://openvpn.net/cloud-vpn} (offering built-in ZT for businesses) due to its open-source nature, widespread adoption, and availability across various platforms.

The VPNaaS logic is implemented as a Helm chart deployed in Kubernetes (K8s), which must interact with other security components in the Security Control Plane, namely with (i) the Keycloak instance in the SDP Controller to process the authentication JWT and (ii) the SDP Gateway to allow the data plane traffic reaches the back-end services.
Once deployed and configured by the Security Intent and Decision engines and the Security Orchestrator, ZTSMs interact across the Security Control and Data Planes in six steps (Fig. \ref{fig:use-case:integrated-view}). First, (1) the SDP client - a human experimenter or non-human entity - obtains its JWT, issued by the SDP Controller; and (2) presents it to its VPNaaS tunnel. Then, (3) VPNaaS authenticates the JWT data validity and timestamp against the SDP Controller. If that succeeds, (4) the data flow from the VPNaaS tunnel can proceed towards the SDP Gateway; which (5) authenticates and authorises the OAuth2-based JWT, using different attributes for each. It is worth noting that each ZTSM performs separate authentication (steps 3, 5) due (i) to its modular nature; and (ii) to separate concerns, as requests to the SDP Gateway could also come from internal, reachable clients.
Finally, (6) the SDP client's request is forwarded throughout the SDP Gateway to reach the protected back-end service.
Further details on ZTSF and ZTSMs (e.g. VPNaaS) are available in \cite{6gbricks_d44} \cite{fernandez_martinez_2024_13626977}.

\subsection{Implementation}

The Helm chart is based on the \textit{cloudnativeapp/openvpn} chart\footnote{https://artifacthub.io/packages/helm/cloudnativeapp/openvpn} and extended to support the project's SSO authentication and networking requirements. 
Its underlying cloud image is also updated with a newer OpenVPN CE (version 2.4.4, released on 25th September 2017), which relies on LibreSSL 2.6.5; and additional dependencies.

\begin{lstlisting}[frame=tb,captionpos=b,belowcaptionskip=0.1em,label=lst:design-implementation:chart:values,caption=VPNaaS' Helm chart values,language=yaml]
...
service:
  type: NodePort
  externalPort: 443
  internalPort: 443
  nodePort: 32000
  hostIP: <k8s_ip>
  ...
keycloak:
  tokenEndpoint: https://<keycloak_realm_oidc_path>/token
  introspectEndpoint: <tokenEndpoint>/introspect
  clientId: <kc_client_id>
  clientSecret: <kc_client_secret>
persistence:
  storageClassName: microk8s-hostpath
  nodeAffinity:
  - <k8s_node_name>
  ...
openvpn:
  ...
  security:
    encryption:
      rsa:
        enabled: false
        diffieHellmanBits: 2048
      ec:
        enabled: true
        curve: "secp384r1"
...
\end{lstlisting}

\begin{lstlisting}[frame=tb,captionpos=b,belowcaptionskip=0.1em,label=lst:design-implementation:chart:auth-iam,caption=AuthN plugin for Keycloak \& OpenVPN,language=python]
import jwt, requests
kctep = "{{ .Values.keycloak.tokenEndpoint }}"
kciep = "{{ .Values.keycloak.introspectEndpoint }}"
cid = "{{ .Values.keycloak.clientId }}"
csec = "{{ .Values.keycloak.clientSecret }}"
tokpt = jwt.decode(
    cred, options={"verify_signature": False}).get("exp")
tok = None
if tokpt is not None:
    tok = cred if tokpt >= int(time.time()) else None
else:
    pld = {"client_id": cid, "client_secret": csec,
           "grant_type": "password",
           "username": user, "password": cred}
    response = requests.post(kctep, data=pld, verify=False)
    if response.status_code == 200:
        tok = response.json().get("access_token")
pld = {"client_id": cid, "client_secret": csec, "token": tok}
if tok is not None:
    response = requests.post(kciep, data=pld, verify=False)
    rc = response.status_code
    return response.json() if rc == 200 else sys.exit(1)
else:
    sys.exit(1)
\end{lstlisting}

This VPNaaS solution relies on logic from two main blocks: (i) a client profile script to generate the .ovpn profile per experimenter, which allows separate exposed ports and dedicated VPN instances; and (ii) the Helm chart files with core logic for key generation and JWT authentication (in its \textit{ConfigMap}) and the persistence, service and deployment details are described (in their respective templates, encoded as K8s resources).

The Helm chart gathers all its configuration in the \textit{values.yaml} file (Listing \ref{lst:design-implementation:chart:values}) and in templates to describe persistence and service needs.
As part of its core logic, its \textit{ConfigMap} incorporates (i) a custom Python authentication script that enables SSO by accepting either the experimenter's OAuth2-based JWT tokens or resorting to \textit{basicauth} otherwise (Listing \ref{lst:design-implementation:chart:auth-iam}) and validating these in the Keycloak IAM; (ii) the OpenVPN configuration (\textit{openvpn.conf} file) with custom values to bind to the authentication script; and (iii) other ancillary scripts for key and certificate generation.

\subsection{Deployment testbed}

The VPNaaS Helm chart was tested across three different K8s environments: a local Minikube setup (v1.33.0), a pre-production vanilla K8s deployment (v1.31.0), and a production MicroK8s deployment (v1.24.17). This multi-platform evaluation ensures robustness and compatibility.

The VPNaaS solution was tested in two different Kubernetes (K8s) environments to assess its robustness, deployment time, and compatibility across platforms: a pre-production vanilla K8s cluster and a production MicroK8s cluster. Each environment had distinct configurations: (i) The pre-production setup used a vanilla Kubernetes v1.31.0 distribution, managed with Helm v3.15.1 on an Ubuntu 22.04.4 LTS system. This setup includes a multi-node cluster with the standard Kubernetes networking stack and uses Persistent Volumes (PV) and Persistent Volume Claims (PVC) for managing persistent data. The goal is to test the core functionality, configuration flexibility, deployment times, and authentication mechanisms before transitioning to the production environment; (ii) The production environment, on the other hand, used MicroK8s v1.24.17 with Helm v3.16.2, also running on Ubuntu 22.04.4 LTS. Unlike the pre-production setup, it used a single-node cluster with MicroK8s' built-in networking stack and relied on PV and PVC for persistence management. This stage was crucial for confirming the solution's performance under conditions closer to real-world production, while also testing the same key parameters validated in the pre-production environment.
\section{Validation}
\label{sec:validation}

The two working modes (RSA or ECDSA) were evaluated in a production environment, using a dedicated HP Enterprise Edge server \cite{6gbricks_d51} with 6 vCPUs at 3GHz, 16 GB RAM and 1 TB disk with a microk8s cluster (version 1.24.17).
As a target KPI, the 6G-BRICKS project mandates establishing the secure end-to-end connections (along with the deployment of the compute continuum services \cite{6gbricks_d23}) in under a minute. Thus, the deployment time is here assessed.
Each VPNaaS Helm chart was deployed using Python's \textit{helmpythonclient} library (version 1.1), which calls a local Helm binary (version 3.16.2).
This runs for one hundred iterations both per working mode (i.e. RSA, ECDSA) and per parameter (i.e. key size, curve).
To ensure clean re-deployments in any testing environment, a new K8s namespace was generated per test, following the format: \textit{6gbricks-vpnaas-\{running\_mode\}-\{param\_iteration\}-\{num\_iteration\}}.
Relevant statistical data were computed on the generated data set, including its Standard Deviation (SD) and Inter-Quartile Range (IQR) to understand the best, worse, average and common cases along with data dispersion.

\begin{table*}[tb]
\vspace{1em}
\begin{minipage}{\columnwidth}
  \centering
  \scalebox{0.95}{
  \begin{tabular}{lcccccc}
    \toprule
    \textbf{Key bits} & \textbf{Min}   & \textbf{Max} & \textbf{Average} & \textbf{Mode}   & \textbf{STD}    & \textbf{IQR}   \\
    \hline \\
    \textbf{512} & 23.383 & 62.327 & 38.024 & 23.383 & 7.163 & 9.000 \\[0.37em]
    \textbf{1024} & 26.278 & 109.618 & 53.288 & 26.278 & 18.019 & 22.218 \\[0.37em]
    \textbf{2048} & 20.935 & 1851.987 & 267.440 & 20.935 & 288.241 & 332.248 \\[0.37em]
    \bottomrule
  \end{tabular}
  }
  \caption{Statistics per VPNaaS-RSA key size (seconds)}
  \label{tab:validation:vpnaas-rsa}
\end{minipage}\hfill
\begin{minipage}{\columnwidth}
    \centering
    \scalebox{0.95}{
    \begin{tabular}{lcccccc}
        \toprule
        \textbf{Curve} & \textbf{Min}   & \textbf{Max} & \textbf{Average} & \textbf{Mode}   & \textbf{STD}    & \textbf{IQR}   \\
        \hline \\
        \textbf{prime256v1} & 24.742 & 57.004 & 38.303 & 24.742 & 6.889 & 9.538 \\
        \textbf{secp384r1} & 14.812 & 36.880 & 25.977 & 14.812 & 3.808 & 2.730 \\
        \textbf{sect409k1} & 18.584 & 35.683 & 26.392 & 18.584 & 3.382 & 3.520 \\
        \textbf{secp521r1} & 18.748 & 66.212 & 36.744 & 18.748 & 7.692 & 10.703 \\
        \bottomrule
    \end{tabular}
    }
    \caption{Statistics per VPNaaS-ECDSA curve (seconds)}
    \label{tab:validation:vpnaas-ecdsa}
  \end{minipage}
\end{table*}

\subsection{VPNaaS using RSA keys}

\begin{figure}[tb]
 \centering
 \begin{subfigure}[b]{0.5\textwidth}
     \centering
     \includegraphics[width=0.95\textwidth]{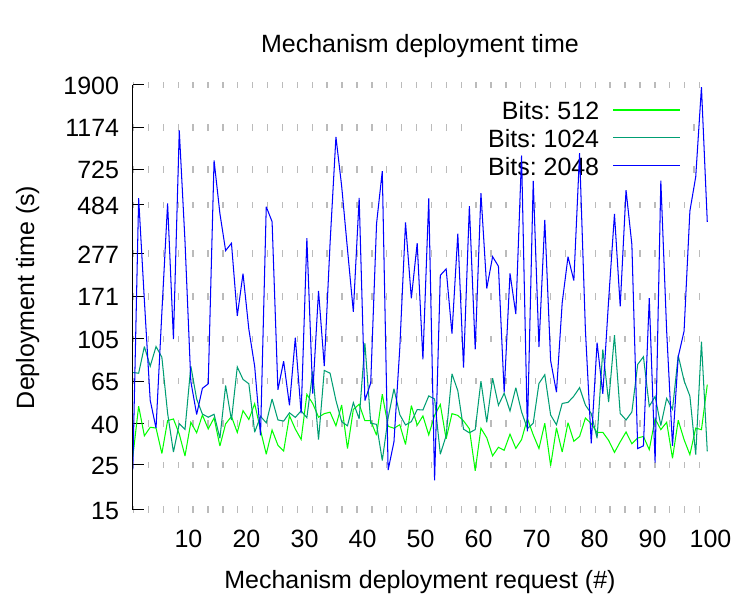}
     \caption{Line plot (logarithmic)}
     \label{fig:validation:vpnaas-rsa:line}
 \end{subfigure}
 \hfill
 \begin{subfigure}[b]{0.5\textwidth}
     \centering
     \includegraphics[width=0.95\textwidth]{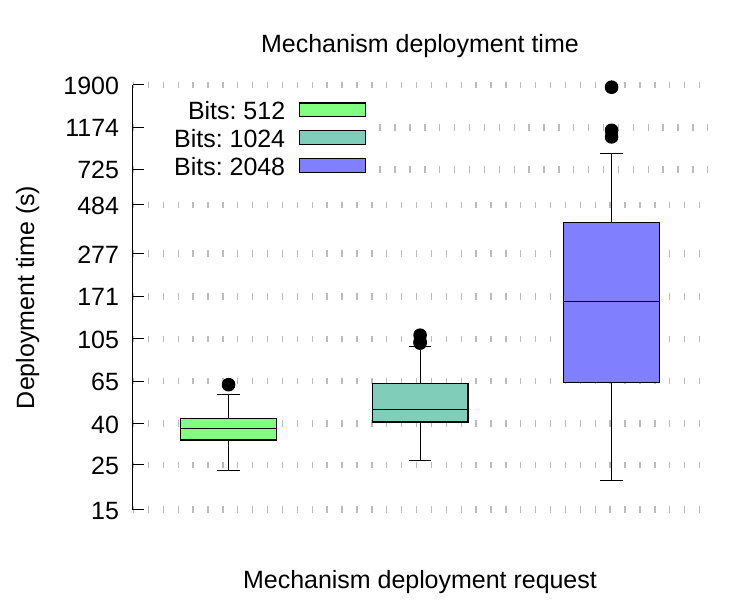}
     \caption{Box plot (logarithmic)}
     \label{fig:validation:vpnaas-rsa:box}
 \end{subfigure}
\caption{VPNaaS deployed with RSA keys}
\label{fig:validation:vpnaas-rsa}
\end{figure}

The first approach for the VPNaaS' tunnel generation was to use RSA keys. This is so since OpenVPN relies on OpenSSL, which uses by default the Diffie-Hellman (DH) key exchange algorithm with 2048-bit RSA keys\footnote{https://docs.openssl.org/master/man1/openssl-req}.
The VPNaaS deployment times in the production environment are shown (running with RSA-generated keys), as line (Fig. \ref{fig:validation:vpnaas-rsa:line}) and box (Fig. \ref{fig:validation:vpnaas-rsa:box}) plots; and in Table \ref{tab:validation:vpnaas-rsa}. These are measured until the service accepts incoming connections to its VPN port.
This figure is plotted with a logarithmic scale (where the Y-axis adopt the results of NumPy's \textit{geomspace} geometric progression) to better represent the results, given the high data variability.
These plots aggregate the deployment times obtained when testing RSA keys of 512, 1024 and 2048 bits.

A singular behaviour was observed during the RSA-based deployment in the production environment, where it shows peak times of around 62, 109 and 1851 seconds yet with mean times of around 38, 53 and 267 seconds for VPNaaS instances with RSA key sizes of 512, 1024 and 2048 bits, respectively (Table \ref{tab:validation:vpnaas-rsa}).
However, when the same VPNaaS instance was deployed in the preproduction environment, it presented peak times of around 9.962 seconds and mean times of 4.708 seconds when using RSA keys of 512, 1024 and 2048 bits, respectively.
Results for tests with 2048 bits are available in \cite{6gbricks_d44}.
The RSA key generation at the edge node in the production environment takes up to the order of 10 to 300 more times than in the pre-production environment.

From the tested key sizes, only RSA keys of at least 2048 bits provide acceptable security (i.e. 112-bit security level \cite{Barker_Roginsky_2019}) until end of 2030 \cite{Barker_Roginsky_2024}, along with RSA keys of up to 3072 bits.
However, their generation may also result in excessive times in some environments and will only grow when switching to larger number of bits; especially considering the 60 seconds threshold for deployment \cite{6gbricks_d23}.

\subsection{VPNaaS using ECDSA keys}

\begin{figure}[tb]
 \centering
 \begin{subfigure}[b]{0.5\textwidth}
     \centering
     \includegraphics[width=0.95\textwidth]{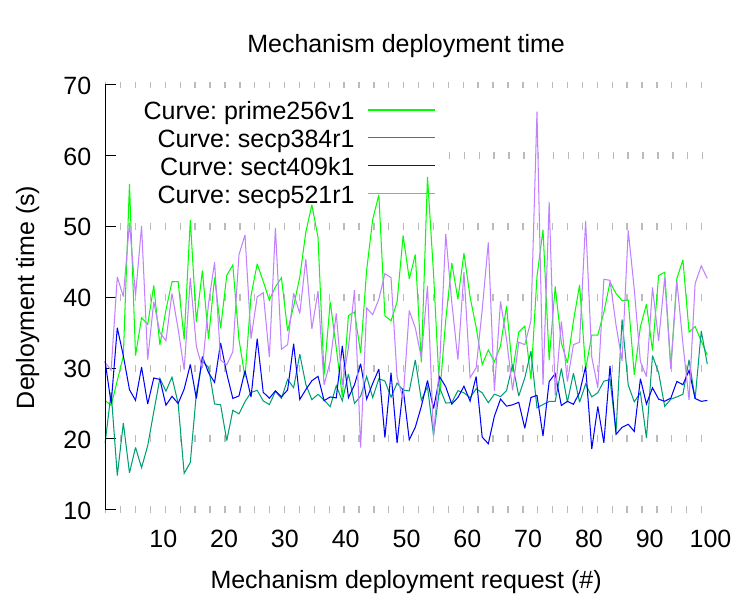}
     \caption{Line plot (linear)}
     \label{fig:validation:vpnaas-ecdsa:line}
 \end{subfigure}
 \hfill
 \begin{subfigure}[b]{0.5\textwidth}
     \centering
     \includegraphics[width=0.95\textwidth]{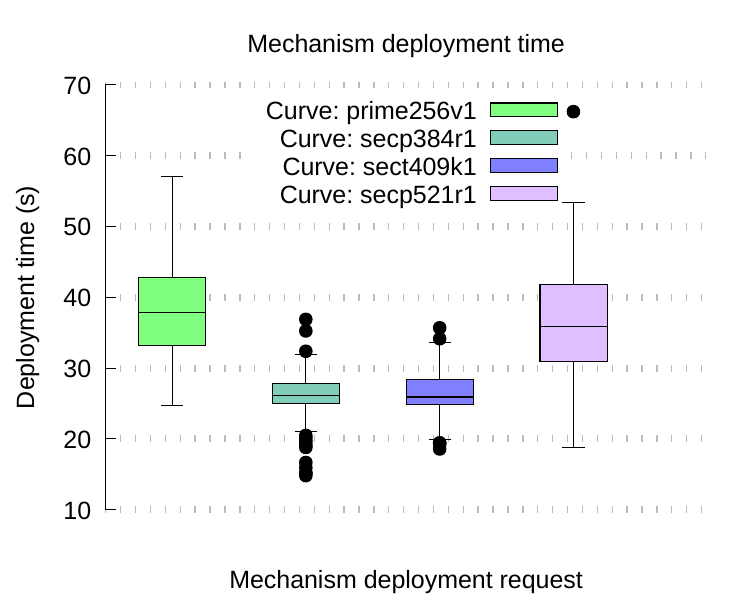}
     \caption{Box plot (linear)}
     \label{fig:validation:vpnaas-ecdsa:box}
 \end{subfigure}
\caption{VPNaaS deployed with ECDSA keys}
\label{fig:validation:vpnaas-ecdsa}
\end{figure}

Given the time threshold and the suggestions on acceptable key sizes for the near future, well-known ECDSA algorithms were investigated to generate the keys, offering less computational expense and better time-bound deployments.

The VPNaaS deployment times in the production environment when running ECDSA keys are pictured as line (Fig. \ref{fig:validation:vpnaas-ecdsa:line}) and box (Fig. \ref{fig:validation:vpnaas-ecdsa:box}) plots, as well as in Table \ref{tab:validation:vpnaas-ecdsa}.
Compared to the deployment results from \cite{Direito_Gomes_Gomes_Aguiar_2023} (min. 285.61 seconds), this solution is up to 11 times faster; even when using EC curves with more bits of security - i.e. 192 compared to 128 (from Curve25519, default in WireGuard\footnote{https://www.wireguard.com/protocol}). This might be explained by more lightweight images, less resources per instance, using containers (rather than VMs), directly interfacing to Kubernetes (instead of using an NFV orchestrator, or NFVO) and also due to simply counting the deployment time until ready, without the NFVO-VNF latency.

As introduced above, any generated key using 112 up to 128 bits of security strength can offer acceptable safety until end of 2030, which translates into 224 to 256 bits for ECDSA keys \cite{Barker_Roginsky_2024}.
Thus, four curves meet such requirements: \textit{prime256v1}, \textit{secp384r1}, \textit{sect409k1} and \textit{secp521r1}; ranging between 128 and 256 -bit security level.
According to the \textit{openssl ecparam -list\_curves} command, all are defined over a bit prime field of the stated number and are NIST/SECG curves except for \textit{prime256v1} (X9.62/SECG curve).

The ECDSA approach has two main benefits: (i) it takes significantly less computing time; and (ii) achieves similar security with reduced lengths, when compared to RSA. ECDSA delivers a better balance between cost and security, as with smaller key sizes it outperforms that of RSA (e.g. a 256-bit ECDSA key is equivalent to a 3072-bit RSA key\footnote{https://www.globalsign.com/en/blog/elliptic-curve-cryptography}).

From the selected curves, \textit{secp384r1} shows better and more reliable performance in the production environment, staying well under the targeted deployment KPI of 1 minute even during peak processing times, as well as having the least variability.
Besides, it also provides equivalent security to that of an RSA key with 7680 bits\footnote{https://www.keylength.com/en/4/}; without incurring into another order of magnitude in computational requirements, as per some of the post-quantum cryptography \cite{Heesch_Adrichem_Attema_Veugen_2019}.
Thus, this is the default parameter for the VPNaaS-ECDSA running mode.
\section{Conclusions}
\label{sec:conclusions}

This work contributed an approach to (i) deliver zero-touch orchestration of multiple VPN instances to serve concurrent, separate secure connections towards separate tenants or clients; (ii) integrate it with well-known IAM tools (i.e. Keycloak) and authentication protocols (i.e. OAuth2); (iii) tailor the VPN deployment to accommodate more secure algorithms (i.e. EC curves) and (iv) select such algorithms and parameters for the key generation, which can also be used to minimise the key generation time used by OpenVPN instances.
Considerations of alternatives and observed anomalies are shared with the reader to provide background to compare with similar cases.
This work is especially suited for environments with long times for computations, e.g. due to resource, entropy or random number generation constraints.

Further enhancements could consider disaggregating the costliest logic (e.g. key generation) to separate instances, thus reducing computational times in more advanced algorithms; while addressing safe key distribution to each VPN instance.
On the other hand, several improvements could be done on the evaluation and adoption of new algorithms (e.g. post-quantum safe) and their parametrisation along with the inclusion of alternative VPN solutions (e.g. WireGuard) or IAMs as a single VPNaaS package.
\section*{Acknowledgment}

This work was funded by the European Union Horizon 2020 research and innovation programme under Grant Agreement no. 101096954 (6G-BRICKS).

\bibliographystyle{IEEEtran}
\bibliography{references}

\end{document}